# Integration of stationary wavelet transform on a dynamic partial reconfiguration 'case study separating preictal gamma oscillations from transitory activities for early build up epileptic seizure

Ridha Jarray[1] and Nawel Jmail[2,3,*] and Abir Hadriche[1,3,4] and Tarek Frikha[5] and Chokri Ben Amar[4]

[1]Université de Sfax, ENIS, REGIM. LAb, Route de Soukra Km 3.5, BP. 1173 – 3038, Sfax, Tunisie.
[2]Université de Sfax, MIRACL. LAb, Sfax, Tunisie.
[3]Université de Sfax, centre de recherche numérique de Sfax, Tunisie.
[4]Université de Gabes, ISIMG, Tunisie
[5]Université de Sfax, ENIS, CES lab, Sfax, Tunisie.

[*]naweljmail@yahoo.fr

**Abstract:**

To define the neural networks responsible of the epileptic seizure, we had to study the electrophysiological signal in a proper way. The early recognition of the seizure build up could also be defined through the time space mapping of the preictal gamma oscillations. The electrophysiological signals present three types of wave: oscillations, spikes, and a mixture of both. Recent studies prove that spikes and oscillations should be separated efficiently to define the accurate neural connectivity for each activity. However retrieving the transitory activity is a sensitive task due to the frequency interfering between the gamma oscillatory and the transitory activities. Many filtering techniques are highlighted to ensure a good separation: reducing false oscillations in the transitory activity and vis versa. However and for a big data set, this separation necessitate a big consumption in time execution, this constraint will be overcome using embedded architecture. The integration of these filtering techniques would also lead to creating instantaneous monitoring of the seizure recognition, build up and neurofeedback devices. We propose here to implement the stationary wavelet transform as a convenient filtering technique to keep only the preictal gamma oscillations on a partial dynamic configuration, then we will use the same architecture to integrate the time space mapping for an early recognition of the build up seizure. We proved a faster recognition of the build up seizure through the non contaminated preictal gamma oscillations time space mapping (about 40 times faster), obtained by the integration of the wavelet transform.

## I. Introduction:

One of the important techniques for diagnosis of neurologic disease is the recording of electrophysiological signal. These signals can be acquired either through electrical potentials (electroencephalography, EEG) or though a magnetic field (magnetoencephalography). Epilepsy is one of the most frequent neurological diseases, and can be either controlled by medication or needs a surgical removal of the epileptic regions. Thus electrophysiology (EEG, MEG) allows delineating the "epileptogenic zone" (EZ): the tissues to be potentially removed by surgery. Following non-invasive brain mapping, it can be decided to implant invasive electrodes in order to better delineate the pathological tissues. These invasive techniques are: intracerebral stereotaxic EEG, electrocorticography (ECoG), and Foramen ovale EEG. The EZ features excessive discharges that may generate seizures; these discharges start from a specific source and propagate, finally implying large cortical areas that form networks [1] [2]. Thus, one has to define signal-based biomarkers of the EZ. In this task, two types of activities are important: the spikes (an activity with high amplitude as defined by Gloor in 1975 [3] and the oscillations. Oishi and colleagues have relied on the localization of epileptic spikes [4] whereas Hiari and his colleague prove the implication of different epileptic oscillations in the beta alpha theta and gamma rhythm [5]. Bragin proved that the high frequency oscillations HFO could be also considered as marker of the EZ [6]–[8].

Urrestarazu, confirmed that the fast oscillations are the best hallmark of the epileptogenic region which are detectable during the build up seizure and accompanied with transient spiking activities. Meanwhile, Bénar and colleagues demonstrated that the oscillatory and transitory activities are difficult to separate as they overlap in the frequency domain. In particular, using simple filters may generate spurious oscillations that are simply the impulse response of the filters [9], thus potentially impacting the result of localization and the definition of the epileptic networks. We proposed in previous work to improve the separation between these activities using several techniques such: Matching pursuit:MP, stationary wavelet transform : SWT and fitting adaptive models ("despiking", i.e. removing transitory activities). We evaluated their performances on simulated data and real electrophysiological signal (IEEG, EEG foramen ovale and MEG) with varying parameters (SNR, rate of overlap, rate of occurrence, frequency, transitory amplitude and width).

These signal processing techniques are expensive in computing time [10]–[12]. In [10], [13], we emphasized also in previous work the importance of defining the space and time presentation of the preictal gamma oscillations in the early detection of the build up seizure. Our goal in the present study is to implement real-time processing of epileptic signals in order to construct in further works a new device for alerting the pharmoco resistant patient of a seizure build up. We implemented the SWT [12] on an embedded system to separate in a convenient way the transitory activities from the preictal gamma oscillatory activities with a minimum time of execution and in a second phase we integrated the time space mapping of the pure preictal gamma oscillations for an early build up seizure recognition. In fact, and now days the embedded systems have invaded the entire field from the industrial to the health and healthcare area [14]. These embedded systems were dedicated also to the biomedical domain to overcome the time execution cost, hence many intelligent architectures were developed to study medical application as the preprocessing of physiological signal ECG MEG , EEG SEEG. The codesign techniques are very relevant in the domain of embedded system, making acquaintance between platforms (SoC, FPGA, ARM, ASIC) and coherent software applied to simplify several biomedical signal preprocessing. The design of embedded systems is an optimal proposal solution and a compromise between several constraints ( cost, chip area, power consumption, and real-time ) [15]. Thus, an embedded system is interpreted as a combination of hardware (processors, sensors, memories) and software (hidden intelligent routines) to achieve in real times many applications as the neurologist decision support with a high degree of security, autonomy and automatic intervention. We propose in this work to integrate the SWT and the time space mapping using a dynamic partial reconfiguration.This manuscript has three sections. The first section is the material, methods and theory calculation (detailing the SWT as a reliable technique for the extraction of the preictal gamma oscillations among the transitory ones, the embedded architecture used for the integration of the SWT filtering techniques, and the morlet wavelet for time space mapping). The second section presented the results and discussion and the third section conclusions and future directions.

## II.    Materials and Methods
### II.1. Materials

The simulated signals and all signal processing was done using the Matlab software (Mathworks,Natick, MA) and the EEGlab toolbox [16]. The implementation of the mixed architecture was embedded on a Xilinx Virtex 5 ML 505 platform, using Matlab and HDL toolbox, as presented in figure 1

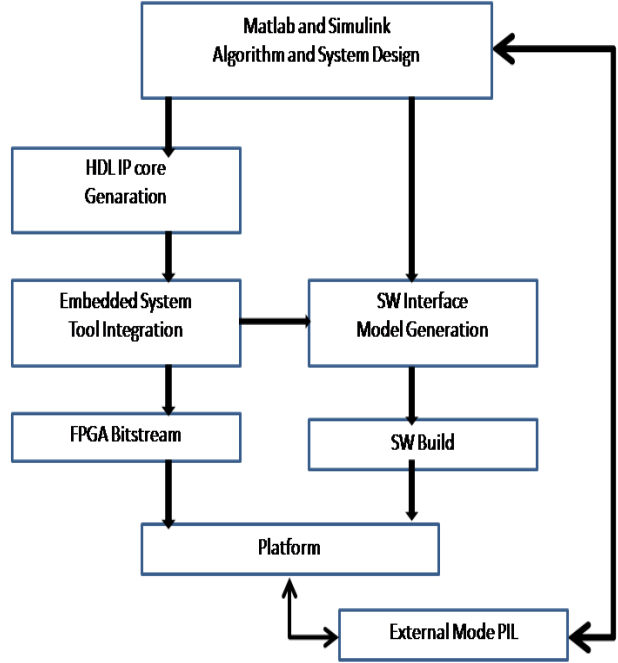

**Fig1**: Materials for Implementation procedure.

**II.1.1. Simulations**

Our simulated data was inspired by an intracererbral EEG obtained from a pharmaco resistant epileptic patient, the sampling frequency was set to 512Hz.We simulated three channels depicting the three occurrence cases between the transitory and the preictal gamma oscillations: activities separated, overlapped and fully overlapped, the preictal gamma oscillations frequency was set to 45Hz for channel 1, 55Hz for channel 2 and 85 for the last channel, we added a colored noise, as in [11]. 200 realizations were generated through the variation of the noise, and the overlap rate using equal steps across the time window of the oscillations.

**II.1.2. Real Signal**

We investigated a presurgical intracerebral EEG signal for a pharmaco resistant subject diagnosed with focal epilepsy. The acquisition and the pre processing phase were applied in the Clinical Neurophysiology Department of the 'La Timone hospital' in Marseille as in [11] studied and validated by an expert neurologist (**Professor Martine Gavaret**).

**II.2. Methods: 'case study: SWT'**

The stationary wavelet transform SWT is a filtering technique that relies on time and scale representation of signals, stationary aspect ensuring time invariance (a translated version of a signal *X* leads to a translation of the transformation), which is effective for representation of spontaneous signals, as well as separation, detection and filtering processes [17].
The SWT was derived from the dynamic wavelet transform DWT (DWT obtained through a step of convolution followed by a decimation) without downsampling the scales. The SWT for a given signal X is the convolution product of X and a tow pass filters (a low one, then a high one) that procures the approximation $C_{j,k}$ and the detail coefficients $W_{j,k}$ consecutively at the j level. These coefficients are obtained by adapting two scale functions $\varphi_k$ and $\Psi_k$ fitted in time (by translation, compression and dilatation) as shown in the following equations (projection of a signal X on these functions)

$$C_{j,k} = \langle s(t)\varphi_k(t)\rangle \qquad W_{j,k} = \langle s(t)\Psi_k(t)\rangle \qquad (1)$$

$$\varphi_k(t) = 2^{-j}\varphi\left(2^{-j}(t-)\right)$$
$$\Psi_k(t) = 2^{-j}\Psi(2^{-j}(t-k)) \qquad (2)$$

After the decomposition step, we have to select the transitory component from the oscillatory one. In this study, our goal is to separate preictal gamma oscillations from the transitory activities in order to perform early recognition of the epileptic build up seizures. More precisely, we aim at a time scale representation of the non contaminated gamma preictal oscillations as in [11]. Thus we used a thresholding step by a rectangular mask with a time spread equal to the length of a gamma oscillatory (200 ms for 45hz, 180 for 55hz and 150ms for 85Hz) and 3 to 2 scales of width (3 scales for 45Hz frequency and 2 scales for the rest) as in [12]. The last step of separation is the reconstruction of the thresholding coefficients (approximation and detail) using the inverse function ISWT on matlab.

**III. Theory / calculation: Embedded architecture to integrate the SWT**

The SWT can be straightly integrated by convolution, which would be also heavy in computation, hence we proposed to use two filters in parallel (a high pass and low pass one), the results of the filtering step will be thresholding to detect the oscillatory activities, this would, ensure a 50% gain in time cost execution [18]. Firstly, we will present the adapted architecture for the SWT integration, then we will describe the chosen platform for this application and finally we will demonstrate the integration of the preictal gamma oscillations detection by the SWT filtering technique.

**III.1. Used Architecture**

We proposed in this work to apply an adaptive architecture based on a partial dynamic reconfiguration [19]. This approach of reconfiguration is predisposed to overcome the complexity of spatial mapping routines [15] with a minimum hardware even for multiple implementations [20].
The integration of a filtering technique requires both local and large scale infrastructures with many constraints that affect the actual separating response. The adopted architecture used in our case (a static and a dynamic bloc) is characterized with a high level of flexibility (further modifications are applied even after the execution), fulfillment and less temporal dissipation during the execution process [21]. We used two accelerators on VHDL for each processor to control the static and dynamic blocs, linked through a Macros Bus. We associated these processors to memory rings to save the separation (Filtering plus thresholding) results, sent back to the processor via FSL, saved in the external memory [22]. These results would be displayed on a terminal screen connected to the FPGA. We used a control unit for the lunch (to generate the accurate address among the memory units) and for the finish (to store the output signal) of the separation routine.
We integrated a Microblaze processor (32-bit embedded microprocessor) to overview the separation datapath units and the results transfers. On the other hand for the dynamic bloc, we used two bits streaming as in [23], with a power PC hardcore processor, that can reach the 150 MHz (The reconfigurable module will operate at the frequency of the static module = 150 MHz), then we added a monocore to handle the multiple iterations of the separation routine [18].

**III.2. Chosen Platform**

The codesign of the hardware/software method was done by adapting the Xilinx Virtex-5ML505 using Embedded Development Kit (EDK) software and the FPGA XC5VLX50T-1FFG1136.
The Xilinx Virtex-5ML505 is featured by large numbers of Input / Output peripherals and an important memory storage that allow a high level of evaluation and smart development of connected, and differentiated systems.

**III.3. Demonstration**

For the simulated data the SWT is repeated 3*200=600 times and for the real signal 96*4=384 (number of channel by the number of realizations /events). Our algorithm could be explained through two rings: the data vector, and the correlation process. Indeed, the data group vector contains 5000 samples for both simulated and real IEEG signal with an amplitude range of [-100 150] µv, which lead to 250 values stored in a multi IO memory [24].
Hence, we have two part the data vector module and the correlation bloc which is preceded in a parallel way, and composed of the convolution of the signal X (5000 samples) with two transfer functions filters $\varphi_k$ and $\Psi_k$, then the thresholding step by a transfer function of a rectangular shape (equation 3), the final results are added and stored in a memory using MATLAB Simulink with Xilinx system generator library for the generation of the VHDL files and the required netlist files.

$$thre_d = \langle C_d, rect \rangle,$$

$$thre_a = \langle C_{ad}, rect \rangle \quad (3)$$

With $thre_d$ are the details coefficients in the rectangular mask adopted for the oscillatory activities and $thre_a$ are the correspondent for the approximate coefficients of the SWT.

**III.4. Time space presentation**

We used the Morlet wavelet transform to assess the spatial temporal map of the preictal pure gamma oscillations. The Morlet wavelet is the result of translation (time shifting) and dilatation (scale shifting) of the mother wavelet function [25]. This mother function is a complex exponential modulated by a Gaussian envelop defined in equation 4:

$$\Psi(t) = \exp(iw_0 t)\exp(-\frac{t^2}{2s^2}) \quad (4)$$

wo denotes the frequency and s is a measure of the spread, this function is shifted in time and scale by the *(a,b)* value as expressed in equation

$$\Psi_{(b,a)}(t) = /a(\exp\left(iw_0\left(\frac{t-b}{a}\right)\right)\exp(-\frac{\left(\frac{t-b}{a}\right)^2}{2s^2}) \quad (5)$$

We proceeded as [11] for the time scale presentation, after the morlet transform, we applied a band pass filter for each data set of the detectible preictal gamma oscillations category ( [40 50] Hz for the 45 Hz oscillations, [50 60] Hz for the 55Hz and [80 90] Hz for the 85Hz gamma oscillations) [26] .Then we applied a convolution between the pass band filtered gamma oscillations and a moving average function with 256 width to smooth the envelope fluctuation [27] of the pure gamma oscillations, finally we applied a normalization step by the lower frequency band as in [28].

*Integrating time space analysis*

For the integration of the time space map, we proceed with the same way of the SWT integration; we used, the same dynamic partial reconfiguration as for embedding the SWT (section III.2.2), manipulated by the FPGA Xilinx Virtex-5ML505. We kept the data vector bloc and we made further modifications for the correlation bloc, in fact we integrated 3 blocs: one for the morlet transform of the gamma oscillations in cascade with the second bloc which treat the smoothing operation and the last ring is dedicated for the normalization operation, the results will be saved in a memory bloc, then delivered to the terminal screen where we can detect the time and space of the build up seizure. See equations below (6)

$$C(nT_e)_{(a,b)} = X(nT_e) * \Psi_{(b,a)}(nTe)$$

$$S(nTe)(a,b) = C(nTe)(a,b) * \sin(2\pi nTe)$$

$$Z(nT_e)_{(a,b)} = S(nT_e)_{(a,b)} * H(nT_e) \quad (6)$$

With X is the original signal, C is morlet transform using the mother morlet function $\Psi$, a and b are the translated and compression value for the morlet transform, S is the smoothing of the C transform via a sinusoidal envelope, H is the transfer function of a band pass filter and Z is the normalized signal of smoothing signal and nTe is the number of samples by the sampling frequency.

## IV. Results and Discussion

Our simulated signal is inspired from a focal epileptic intracerabral EEG signal, it presents the preicatal and the onset seizure, three channels with a mixture of transitory and preictal gamma activities for three different frequency range (45 55 and 85 Hz) depicted in figure 2 with a time course of the real IEEG signal depicting the same mixture of activities.

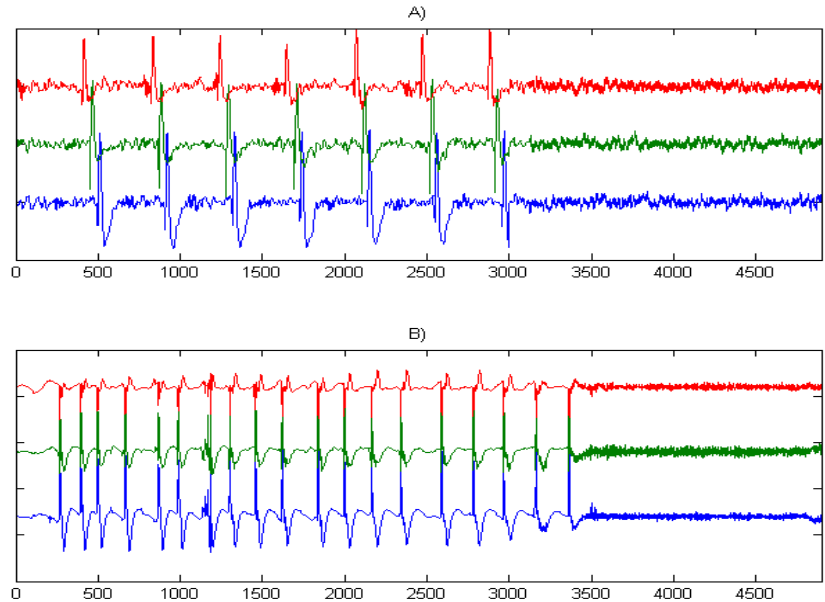

**Fig2**:A) One realization of simulated data, each channel reproduce the 3 cases of overlap between transient and oscillations activities for 3 range of frequency 45, 55, and 85 Hz. B) three channel of real IEEG signal with transitory, preictal gamma oscillations and seizure with different frequency range between 45 and 85 Hz.

The figure 3 illustrates the processing steps for the separation between transitory and preictal gamma oscillatory activities using the SWT and the rectangular mask for the thresholding phase then, the time space mapping of the pure preictal gamma oscillations.

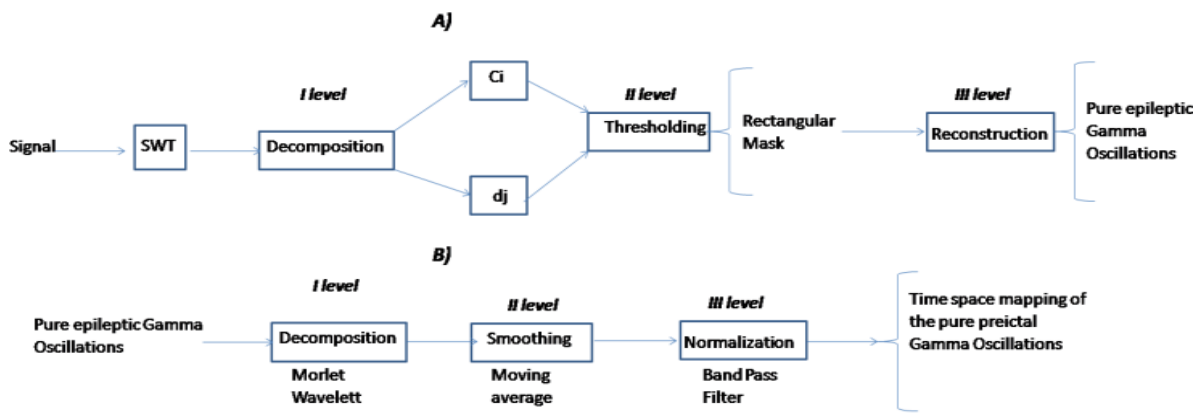

**Fig 3**: Separating between transitory and gamma oscillatory by SWT filtering technique and time space mapping of the pure preictal gamma oscillations.

Firstly, and for the adequate separation between transitory and oscillatory activities, we proceeded on the simulated data, we applied the SWT for each channel, and then we applied the thresholding step and finally, the reconstruction of the pure preictal gamma oscillations, we proceeded in the same way for the real IEEG signal.

For the time space mapping, we did illustrate the energy of all the implied channels (for both simulated data and real IEEG signal) before and after the detection of the preictal gamma oscillation to define the build up seizure after retrieving all the transitory activities through three steps (decomposition, smoothing and normalization).

In figure 4 we depict the results of the reconstruction of the gamma oscillations among the original preictal signal, illustrating the shape of mask used for the detection of the transitory activity and the gamma oscillations (we adopted two masks a rectangular one for oscillation and a pyramidal mask for transitory activity) and the variation of the rectangular shape (length and width) with the increasing of the frequency range from 45 Hz to 85 Hz.

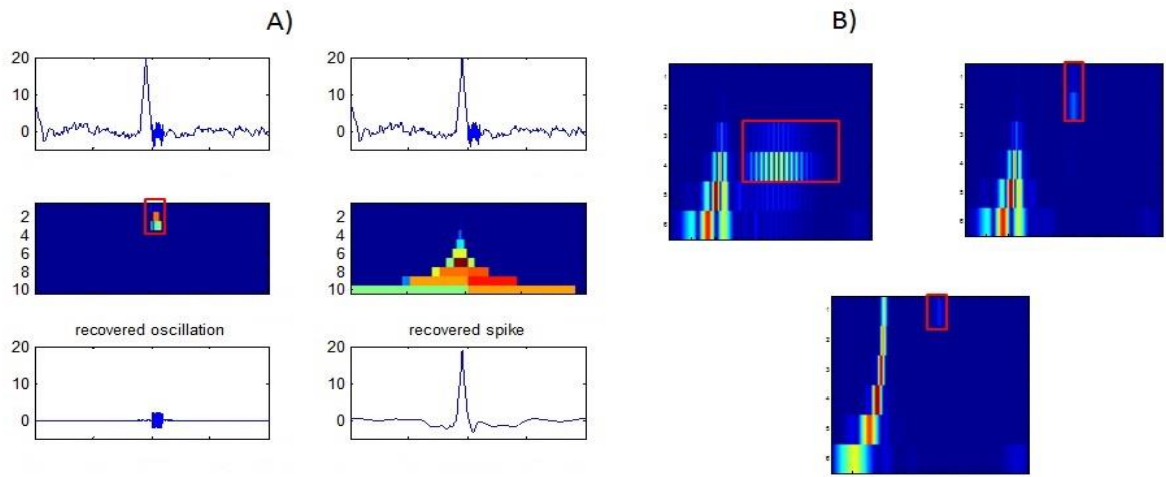

**Fig 4**: A) line 1: mixture of transitory and gamma oscillation of 45hz in full overlap, line 2 : the masks used for threshollding the transitory activity among the oscillatory one, line 3 the reconstructed gamma oscillations and transitory activity after using the SWT and thresholding process. B) the approximation coefficients of the SWT for three range of gamma oscillations 45, 55, and 85 Hz with a rectangular mask to reconstruct only the preictal gamma oscillations components.

We demonstrate in figure 5 the automatic detection of the preictal gamma oscillations of 55 Hz from the intracerebral EEG signals. We illustrate the original signal, and the recovered preictal gamma oscillatory activities for the preictal and seizure too.

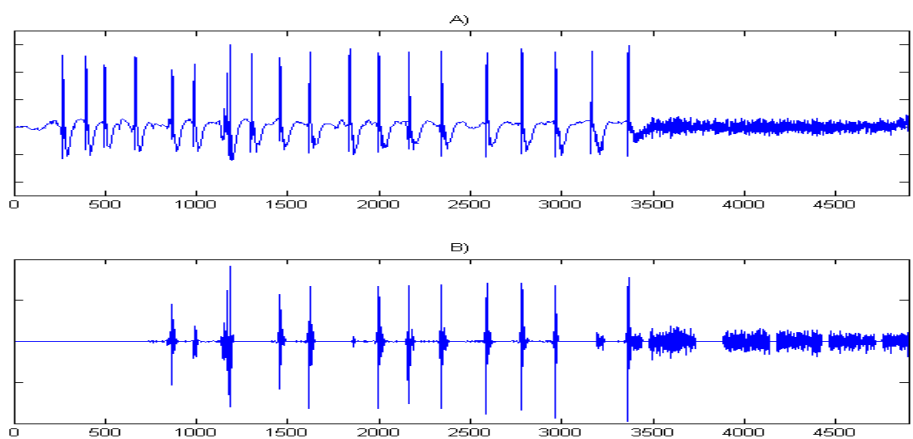

**Fig 5**: A) : IEEG signal presenting a mixture between transitory and preictal gamma oscillatory activities. B) the detected preictal gamma oscillations by the SWT.

We depicted the spatio-temporal maps in figure 6, for the preictal gamma oscillation of 85Hz, for all channels of our IEEG signal used in this work, normalized by a pass band filter of 10–15 Hz band and smoothed with a moving average of 256 samples. The seizure build up is clearer for the signal with only preictal gamma oscillations (no transitory activities) which proves that the large energy of spikes would walnut this phase (preictal). Several channels present energy in high frequency band during the seizure onset, which prove that interictal oscillations may be considered as a supplement biomarker of the epileptogenic zone.

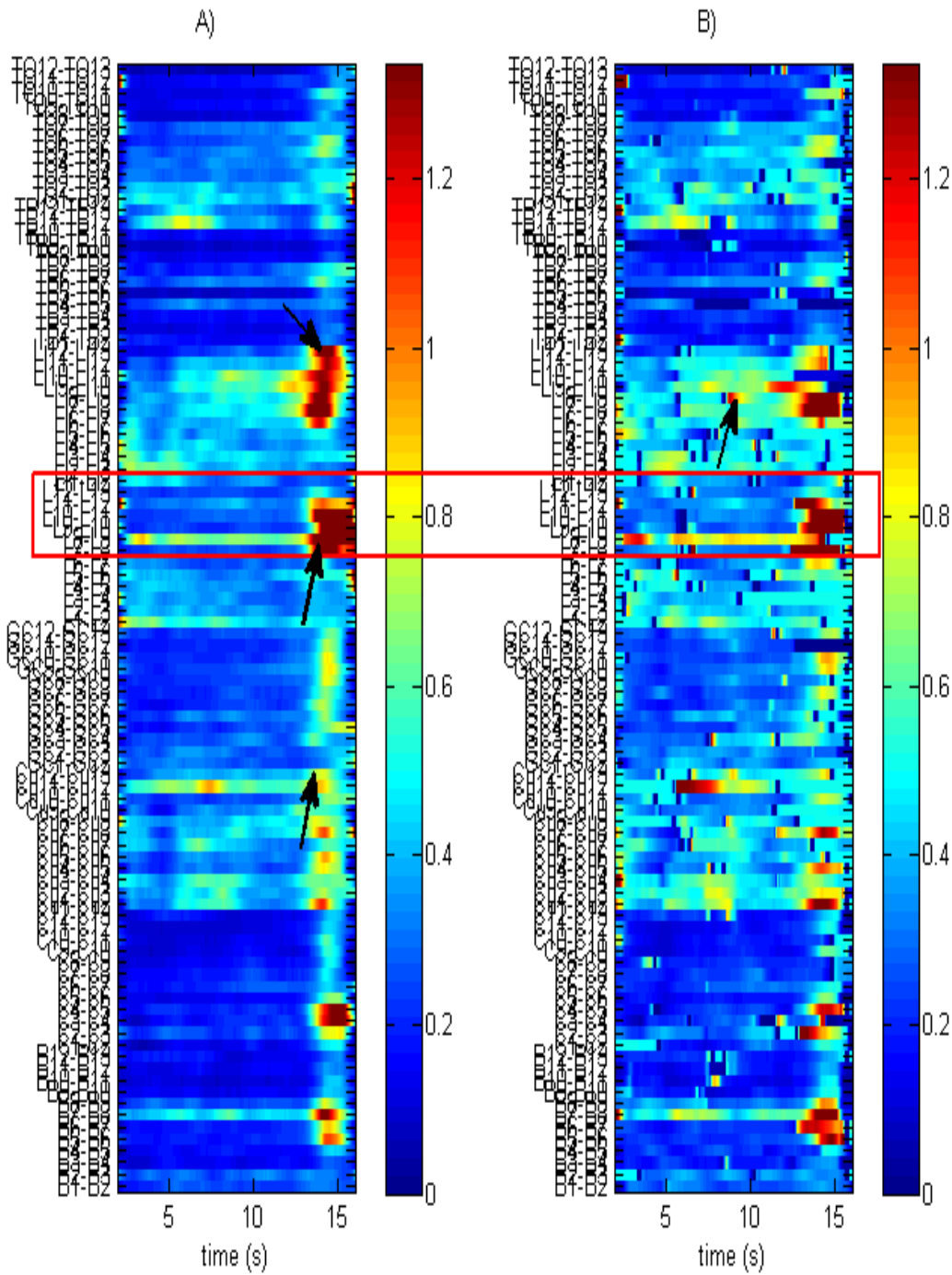

**Fig 6**: Spatio-temporal maps of IEEG signals (channels *x* time), for preictal gamma oscillations of 85Hz, A) IEEG time space map: no clear pattern in the preictal period and seizure is clear in several regions (L7, L8 to L12). B) IEEG time space map after retrieving the transitory activities 'Despiked signals By SWT': much more clear 'build-up' of oscillatory activity and seizure seen in channels L8, L9, L10 and L11.

Figure 7 depicts the dynamic partial reconfiguration for the integration of the separating technique (SWT plus thresholding). The partial reconfiguration contains two blocs a static and a dynamic one, connected to a processor (Microblaze) and an external memory unit with a Macros Bus.

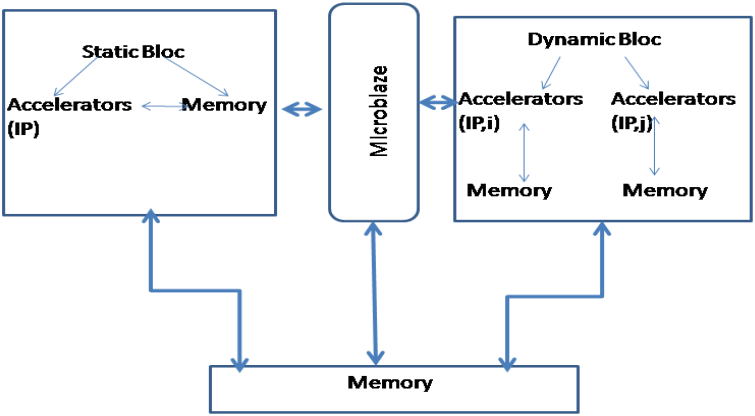

**Fig7**: the partial reconfiguration composed of tow bloc a static and a dynamic one connected to the processor and to an external memory. The static bloc is handled by accelerators and memory for instantaneous saving and the dynamic bloc is composed by two accelerators and memory units.

Figure 8 depicts the adopted architecture for the integration of the separating technique (SWT plus thresholding), the entire adopted architecture is composed of a data vector bloc and a correlation bloc. The correlation bloc is the algorithm hardware bloc, the most important bloc of the adopted architecture featured by the independence towards the rest blocs. It translates the three convolution operations to keep only the oscillatory components from an original Signal. Thus we used 3 sets of addition and multiplication operators each set is dedicated for one convolution operation (down sampling the scale functions, decomposition and thresholding). The multiplication and addition operations are set in a parallel way: the first operation is for the down sampling of the scale function, the second is dedicated for the convolution between the signal (simulated and real IEEG) and the transfer functions of the two filters ( scale function) which results in tow types of coefficients (approximation and detail coefficients), and the last operation is used for the thresholding process (convolution between the SWT coefficients and a rectangular Mask).These Blocs convolution operation ( a set of addition and multiplication operation ) will be executed in parallel since we integrated 2 accelerators as in [23]. The results would be stored straightly in two separated internal memory (Msd/Msa for scaling function or filters, Md/Ma: approximation and detail coefficient for the SWT and Mat/Mdt for thresholding the approximation and detail coefficient by a rectangular mask). Then the oscillatory components (approximation and detail thresholding) will be added and saved in the external Memory bloc. Thus the external memory unit will store only the preictal gamma oscillatory components obtained either from simulated data or real IEEG signal which saved in the data vector bloc to be ruled by the adopted architecture used above.

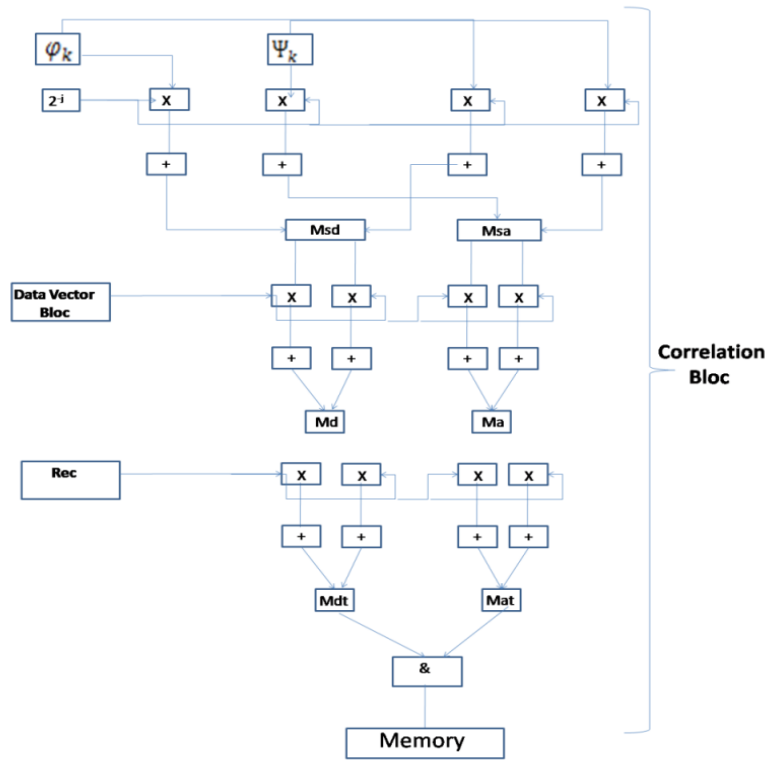

Fig8: the adopted architecture for the integration of the separating technique is composed of two primordial bloc: the data vector bloc to receive and save the input signal (simulated/IEEG) and the correlation bloc which contain 3 operations: multiplication/addition, 3internal memories bloc and an external memory unit to restore the approximation and the detail coefficient after thresholding by a rectangular mask.

In fact, we profiled our algorithm for preictal gamma oscillation detection using in first place no accelerators then after the integration of 2 accelerators. In table 1, and after the stimulation test in Xilinx ISE, we collect the results of the time consumption of running the algorithm of the preictal gamma oscillation detection using the SW transform. Firstly we tested the time consumption of our embedded algorithm using no accelerators then after integrating 2 accelerators. In fact the absence of accelerators has alleviated the time execution per tics but is still heavy in consumption; the implementation of 2 accelerators induced almost 22 tics of time gain.

Table1: Time consumption of the SWT integration

| SWT Integration | Time per Tics |
|---|---|
| Software | 420 |
| No Accelerator | 45 |
| 2 Accelerators | 23 |

Figure 9 illustrates the integration of the time space mapping of the preictal gamma oscillations, the adopted platform is composed by three blocs each one treat the three sets of convolution (Morlet transform, smoothing and normalization) each bloc is composed by an addition and multiplication operators, after each bloc we used an internal memory unit to store the results (Mw for wavelet transform, Ms for smoothing and Mn for normalization). The input Ψ, H, sin and Z denote the transfer functions used for our algorithm of mapping. After compiling the three convolutions we used a comparator to detect the maximum energy through the captors (channels of the original data). The output result would be stored in an external memory unit then mapped on a screen using the comparator.

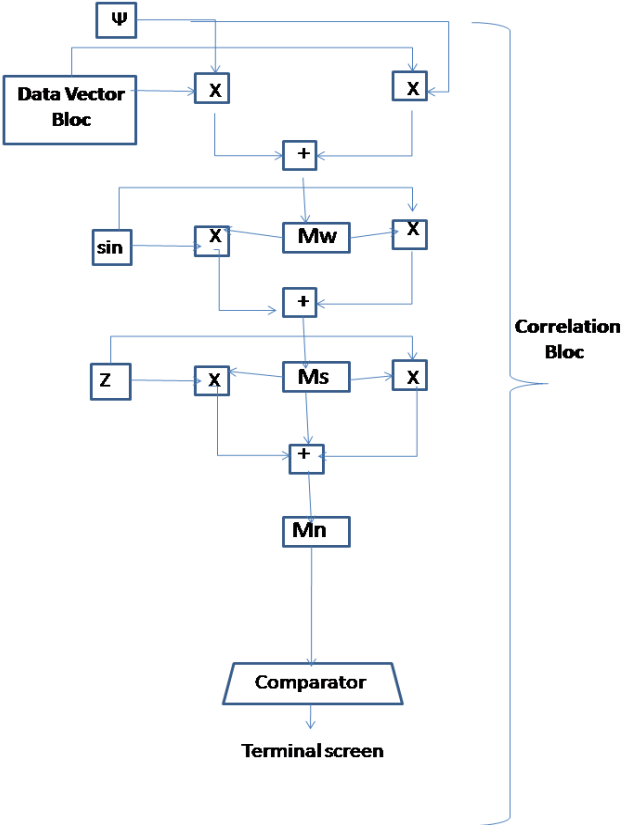

Fig 9: the integration of time space mapping is composed of a data vector (that stores the preictal gamma oscillatory activities), the correlation bloc consists of three convolution sets each one is presented by an addition and multiplication operators the saved data will be incorporated into a comparator to display the space and time with highest energy to predict the build up seizure.

The integration of the time space mapping was done adopting the same architecture for the integration of the SWT: a dynamic partial reconfiguration on the Xilinx platform, using a data vector bloc to access the input signal (original signal/despikied signal (no transitory activities) and a correlation bloc with three multiplication/addition rings , 3 memory units and an external memory bloc to store the energetic component of the morlet transform of the preictal gamma oscillations).

In table 2, we collect the results of the time consumption for embedding the spatio temporal mapping of the preictal gamma oscillations , as demonstrated in the first table (without accelerators, then implementing 2 accelerators), it 's clear that using 2 accelerators made about 20 tics of time gain.

Table2: Time consumption of the spatio temporal mapping

| Embedding spatio-temporal mapping | Time per Tics |
|---|---|
| **Software** | 320 |
| **No Accelerator** | 35 |
| **2 Accelerators** | 16 |

In Table 3, we represent the time consumption in slices/ flip flop slices and BRAM for the entire integration procedure from the detection of the preictal gamma oscillations to the space time mapping using the Xilinx ML 505 Platform Studio to define the exploitation rate of the proposed embedding system for the pre processing algorithm of the pure preictal gamma oscillation time space mapping .

Table 3: The integration of the entire procedure of the space time mapping of the preictal gamma oscillations per Logic uses.

| Logic Manipulation | Available | Hold | Rate |
|---|---|---|---|
| Slices | 7616 | 11200 | 68% |
| Flip Flops Slices | 28224 | 44800 | 63% |
| BRAMS | 144.5 | 228 | 63.5% |

The rate of logic uses is about 70% of the available capacity which imply a good exploitation of the proposed embedded architecture for the integration of the early recognition of the build up seizure.

Finally, we collect in table 4 the energy consumption of the embedding system used for the integration of the preictal gamma oscillations time space mapping. The energy required for this procedure is acceptable for embedding systems and the Xilinx platform, even after adding two accelerators, a simple battery is sufficient to supply power to our used system.

Table4: The energy consumption of the integration of the preictal gamma oscillations in the time space mapping.

| Integrating Procedure | Energy consumption |
|---|---|
| Absence of Accelerator | 1480 mW |
| Two Accelerator | 3000 mW |

Although the energy consumption is about twice the need of integrating our algorithm with no accelerator, but this consumption is still low and acceptable in the integration field of a preprocessing file still the benefit of adding two accelerators have an important impact on the time gain consumption which prove that using two accelerators is very efficient in the proposed embedded system.

V. **Conclusion and Perspectives**

The early recognition of the epileptic seizure is an important task in the diagnosis of epilepsy and especially for the pharmaco resistant patients. The analysis of the electrophysiological signal (EEG, MEG, EEGFo , IEEG, ECoG) is fundamental in this field; however, these signals show a mixture of activities (transient and oscillatory waves) which could imply different cortical regions and different sources. In previous work, Bénar and his collaborators emphasized the sensitivity of the separation between these activities and suggested reliable filtering techniques ,evaluated against several constraints [1], [9]–[12]. Among these techniques the SWT was shown to be a convenient tool for separation of signals. In this study we applied the SWT to "despikify" simulated and intracerebral EEG signal by keeping only the preictal gamma activities. Despiking was firstly applied and validated on simulated data inspired from intracerebral EEG epileptic signal, then on real IEEG signal with focal epilepsy. The time scale study of these despiked signals lead to a better characterization of the time and the localizations of regions responsible of the excessive discharges and hence of the build up seizure ( this was highlighted in previous work [11], [13]). However these pre processing steps are heavy in calculation which called the immediate integrated systems. The integration of the SWT was proposed using a dynamic partial reconfiguration based on two blocs: a static and a dynamic one, which improved drastically the time execution. In fact, we maintained the same results of separation between transitory and preictal gamma oscillations with almost 40 times faster than the non integrated filtering technique. Our embedded system proposed is based on two important units: the data vector bloc and the correlation bloc. In fact, the correlation Bloc is the most important ring in the adopted architecture, it s composed of 3 memory units (to store results of convolution) and 3 sets of addition and multiplication operators that profile the convolution operation.  In a second step we implemented the time space mapping of the non contaminated preictal gamma oscillations using the same adopted architecture and reconfiguration as the ones applied for the SWT integration. We employed also an external memory bloc to save the energy of the preictal gamma oscillatory activities then mapped in an external screen lineked to the FPGA. Two accelerators were managed for each integration process to ameliorate the time consumption and to increase the logic manipulation of the adopted architecture that reaches almost 70% of the available capacity and 20 tics gain in the computation costs. The energy consumption is convenient for the Xilinx platform, even when we added two accelerators; a simple buttery can handle the power needs of our proposed system.  These results could help the neurologists for an early seizure recognition and detection, and could be also implemented in a portable device for epileptic subjects to alarm them about the build up of an epileptic seizure and the need to take the necessary precautions to reduce risks.

We proposed in the next work, firstly to compare the robustness of despikification by SWT against SVD developed in [11] and to propose an intelligent architecture for the integration of the SVD too. Secondly to proceed on the computation of networks connectivity of the preictal gamma oscillations to define the accurate sources responsible of the excessive discharges leading to build up a seizure, all these routines should be in second time integrated to propose a neurofeedbak and monitoring devices for neurologist support decision, to reduce the immediate damages of an onset seizure, and overcome several health care problems.

**Conflict of Interest:** The authors declare that they have no conflict of interest.